# *Correlation between objective and subjective assessment of noise barriers*

J. Redondo[1*]; M.P. Peiró-Torres[2]; C. Llinares[3]; J.M. Bravo[4]; A. Pereira[5], P. Amado-Mendes[5]

(1) Universitat Politècnica de València. Instituto de Investigación para la Gestión Integrada de zonas Costeras, Paranimf 1, Grao de Gandia, Valencia (Spain); fredondo@fis.upv.es
(2) BECSA, S.A.U. Ciudad del Transporte II. C/ Grecia, 31, Castellón (Spain); mppeiro@becsa.es
(3) Instituto de Investigación en Innovación en Bioingeniería (i3B), Universitat Politècnica de València, Camino de Vera s/n, Valencia (Spain); cllinare@omp.upv.es
(4) Universitat Politècnica de València. Centro de Tecnologías Físicas, Acústica, Materiales y Astrofísica, División acústica. Camino de Vera s/n, Valencia (Spain); jobrapla@fis.upv.es
(5) University of Coimbra, ISISE, Department of Civil Engineering, Portugal ; pamendes@dec.uc.pt; apereira@dec.uc.pt

## Abstract

There are several international standards that define the way to evaluate the attenuation capacity of noise reducing devices, by single-number quantities representing airborne sound insulation and insertion loss. These two single-value ratings define the quality and performance of acoustic barriers, the former being related to intrinsic and the latter to both intrinsic and extrinsic acoustic characteristics of the devices. However, not many studies can be found on whether these objective parameters correlate to the perception of annoyance reduction.

The aim of the present work is to analyze the adequacy of these objective ratings to indicate the performance of noise barriers, by comparing their values with the perception of annoyance reduction.

For this purpose, ninety individuals of two different nationalities (Spanish and Portuguese) were asked to rate the perceived annoyance reduction in a listening experimental test, in which they were exposed, under controlled conditions, to several environmental noises and acoustic screened stimuli simulated by audio filters.

The obtained results show a high correlation between objective ratings and subjective annoyance perception, with a better correlation being observed for insertion loss single-number parameter than for the airborne sound insulation single-number rating.

SCAS: Sonic Crystal Acoustic Screens
WHO: World Health Organization
BG: Band Gaps
PAR: Perceived Annoyance Reduction
SPL: Sound Pressure Level

Furthermore, significant differences were found depending on the gender and nationality of the respondents. The results, from this ongoing research work, may be of great interest for future acoustic barriers design.

**Keywords:** Noise barrier, Perceptual assessment, Annoyance reduction, Airborne sound insulation, Insertion loss, Attenuation sound environment.

**1. Introduction**

During the last decades, the increasing number of vehicles in urban zones has lead to excessive environmental noise pollution, which is mostly caused by road, railway and aircraft traffics. Among these, the most significant source of noise is road traffic [1], exposure to which far exceeds rail and aircraft sources combined [2]. In fact, in urban areas, road traffic is thought to account for 80% of all noise pollution [3]. It is therefore very important to achieve lower sound levels from road traffic in the process of planning urban environments [4].

Several studies have demonstrated that environmental noise is an important public health issue. The recent report "Environmental Noise Guidelines for the European Region", from the World Health Organization (WHO) [5], provides the current state of knowledge about the non-auditory effects due to environmental noise on the population health. Noise affects cardiovascular diseases such as ischemic heart disease, hypertension or strokes, cognitive development, sleep disturbance and variables that have an adverse effect on birth and even other variables related to the decrease in quality of life and metabolic diseases. In Europe, environmental noise is assumed as an important public health issue, being among the top environmental risks to health. Its negative impacts on human health and well-being are a growing concern among both the general public and policy-makers in Europe [6]. Being aware of this problem the WHO has, in 2018, updated the guidelines for the European Region, in order to protect human health from exposure to environmental noise [5].

In general, noise propagation can be controlled in three different ways: (i) reducing noise generation near the source; (ii) controlling noise propagation from source to receiver, and (iii) taking measures near noise reception. In the former case, acting near the noise source corresponds, for example, to reducing engines sound power, to reducing vehicles speed or to adopting noise absorbing pavements or more silent tyres. On the other hand, regarding noise reception, sound insulation of buildings and buildings facades has to be considered, although being a complex and expensive task [7].

The most commonly employed solution to reduce road traffic noise is the use of noise barriers, which mitigate noise by placing an obstacle between noise emission and reception. Some of the emitted sound energy is reflected or dispersed towards the source, some of the energy is absorbed and dissipated by the barrier material, and some energy arrives at the receiver, being diffracted from the edges of the barrier or being transmitted through this type of noise reducing device [7].

Usually, a noise barrier is a solid continuous, opaque and appropriately dense construction. The effectiveness of these devices in reducing noise is related to several factors, such as the relative position between the noise emitter and receiver, the barrier's height, length, thickness or its geometric design, the presence of top diffusive devices, or ground cover in the vicinity of the barrier [8, 9].

There are different types of barriers, namely: simple reflecting barriers, absorbing/diffusive barriers which have absorbing materials on the side facing the noise source; angled barriers which reflect sound away from the receiver with a specially designed geometry and diffusive top section; and covering barriers, such as galleries or tunnels, that offer significant noise reduction. Noise barriers can be made of a variety of materials, such as glass or plastic/acrylic thin elements, masonry blocks, pre-cast concrete elements, perforated steel or aluminum, and they may also incorporate recycled materials [10-11].

The use of classical sound barriers in urban areas can have considerable disadvantages. One of these is clearly related to the reflection of sound energy that occurs on the surface of the traditional barrier and can significantly affect receivers on the same side of the sound source. This is mitigated in the case of Sonic Crystal Acoustic Screens (SCAS) by producing a reflection with a higher diffusion index than the traditional barrier [12]. Undoubtedly, the sound absorption of the acoustic barrier is a very important factor in the design and rethinking of the barriers. However, this work will not take this into account as its main objective is the evaluation of the perceived annoyance of the noise energy passing through the barrier (either through it or by edge diffraction).

Other factors associated to this new barrier technology (SCAS) are related to the reduction of the necessary foundations due to the decrease in the effect of wind load since the barrier surfaces are much more permeable. Permeability also affects the concentration of pollutants and temperature near the ground, as traditional barriers prevent the passage of air, they generate an increase in temperature in the environment and they constitute a physical limit that affects the natural dispersion of pollutants. However, SCAS, being permeable, do not produce such effects.

In addition, the length and height of the large opaque panels have a strong effect in relation to blocking the field of vision of citizens and reducing natural light, so they have a significant impact on the urban landscape and provide physical isolation of acoustically protected areas [13]. In general, it could be said that traditional acoustic barriers have not evolved much in recent years from a technological point of view; in fact, they are still non-tunable acoustic systems that act in the same way regardless of the spectral characteristics of the noise, and are often inefficient at low frequencies [14,15].

In the last two decades, new devices to provide noise reduction on urban environments have been developed based on disruptive concepts. Among these are the SCAS which consist of structures built by periodic arrays of cylindrical acoustic scatterers, separated by a predetermined lattice constant [16]. In this type of structures, sound attenuation is provided for certain frequency bands, by activating a noise control mechanism based on the Bragg interferences due to a multiple scattering process [17]. The range and position of these frequency forbidden bands, also called Band Gaps (BG), can be designed by changing the geometrical properties of the arrays of scatterers [18, 19]. Nowadays, the use of SCAS is at an intermediate state between the basic research of concepts and physical properties, and their industrial production and widespread use as noise reducing devices.

The idea of using arrays of scatterers for sound attenuation was born from the artistic sculpture by Eusebio Sempere, with a periodic arrangement of steel tubes installed in the gardens of the Juan March Foundation in Madrid (Spain). Despite the improvement of aesthetic characteristics of these devices, they also enable some visual continuity to the urban landscape, since a complete interference of the optical line between emitter and receiver is not produced through the discrete scatterers forming the SCAS. Another advantage of this type of sound barrier is its permeability to wind, substantially reducing the effects of turbulence and the forces exerted on the ground, and allowing for lighter foundations during on site implementation.

In order to increase the sound attenuation provided by the SCAS, it is also possible to add other noise control mechanisms, such as absorption or resonance [20], in addition to the BG effect, allowing acoustic tuning [21] in the design of the noise barrier for each noise spectra [22]. The use of SCAS has also been explored because of its high environmental sustainability, when logs from forest thinning operations have been proposed for traffic noise abatement [23]. As a matter of fact, there is ongoing research work envisaging the increase of the range of frequency bands that can be attenuated by SACS in traffic noise mitigation, as has been described in the revision work by Fredianelli et al. [24] or as it can had seen in the recent paper by Gulia and Gupta [25], where variations of systems based on sonic crystals have studied with the aim of extending their performance bandwidth and improving their noise attenuation efficiency.

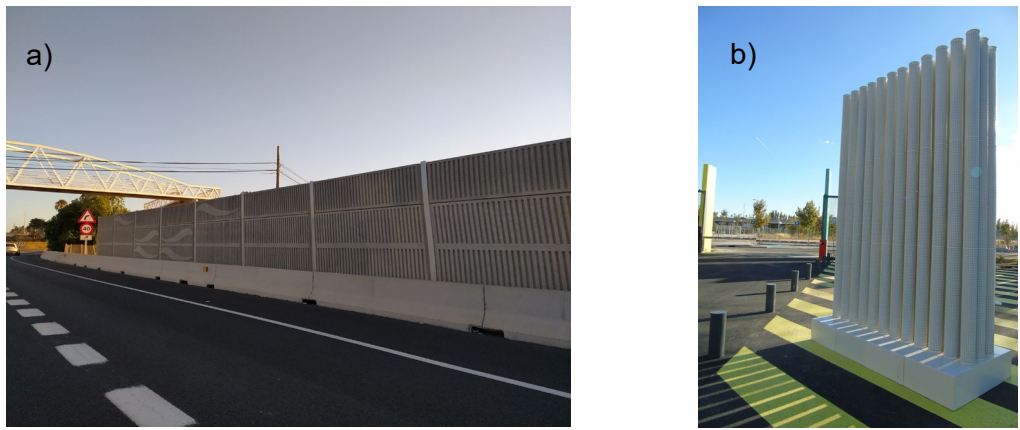

Figure 1. a) photograph of a traditional continuous noise barrier; b) photograph of a noise barrier based on sonyc crystals concept.

The effectiveness of a sound barrier, in terms of the sound attenuation provided, can be expressed by the following parameters: two intrinsic parameters, $DL_R$ and $DL_{SI}$, which characterize the attenuation of sound propagation, and an extrinsic parameter, $D_{IL}$, which takes into account the physical characteristics of the sound barrier, its height, thickness and the position of the barrier in relation to sound emission and reception. These parameters are given by:

- *$DL_R$*, which refers to a single-number rating of airborne sound insulation for devices designed to reduce road traffic noise under diffuse sound field conditions in the laboratory [26];
- *$DL_{SI}$*, which refers to a single-number rating of in situ airborne sound insulation of the noise reducing device for free field sound conditions [27];
- *$D_{IL}$*, which corresponds to an insertion loss value, that is evaluated by the difference, in decibels, in sound pressure levels registered at a specific receiver position before and after the installation of an outdoor noise barrier [28].

Some published works can be found that describe methodologies for in-situ measurement of sound reflection and airborne noise insulation characteristics of acoustic barriers [29, 30, 31]. Some works studied how tonal sounds are perceived by the human ear or tried to understand human auditory comfort through psychoacoustic studies [32, 33]. However, few papers have analized whether the objective value of sound attenuation of such devices corresponds to the subjective perception of the population. Some of these studies have been carried out indoors, by comparing traffic noise façade insulation linked to reported noise annoyance [34], by consulting a sample of the population in order to relate the users' opinion to airborne sound insulation of constructions [35] or to impact sound insulation over a concrete floor with different coverings [36]. Also some study have studied the intersensory perceptions of noise barrier performance in terms of the noise reduction combined with visual impressions [37] but in no case it has been studied whether there are significant differences that affect of the users features like gender or nationality.

The objective of the present study is to analyze the suitability of quantifying the performance of noise barriers by comparing the above mentioned objective ratings to indicators with the subjective response of the people surveyed in listening tests.

For this purpose, the acoustic characteristics of the devices performances, as well as those of the subjects, will be taken into account. Firstly, the results of the analysis of the surveyed Perceived Annoyance Reduction (PAR) and its correlation with the objective parameters of the sound barriers, intrinsic and extrinsic, will be analyzed. Secondly, that correlation is studied according to the characteristics of the population interviewed and different types of emittied sound.

The observed limits of these objective parameters are investigated, trying to address the following questions: Does it make sense to continuously improve the results of target objective indices? Does the human ear respond in a linear way to this improvement? Is there a possible saturation limit beyond which the improvement of the barriers is no longer useful, since the human ear does not perceive this improvement? In fact, if a saturation limit can be set, it would be very interesting to be able to define the range in which this saturation occurs, as this would correspond to the limit for improving the objective sound attenuation of these reducing noise devices.

In short, this work aims to address the psychoacoustic research field, by studying the human perception of noise under the influence of noise reducing devices, such as acoustic barriers, that are used to attenuate road traffic noise.

## 2. Testing methodology

This section will present the design and methodology of the subjective experiment that was carried out. Then, the type of traffic noise and the adopted signal processing procedure to induce different degrees of sound attenuation will be explained. Finally, details will be given regarding the procedure followed for the collection of information through listening tests.

The listening surveys present an international character, and they were carried out at both the offices of the Department of Civil Engineering of the Universidade de Coimbra (in Portugal), and at the offices of the Sonic Crystal Technology Research Group at the Universitat Politècnica de València (in Spain).

### 2.1. Sample of participants.

To develop the present study a sample of 90 people, from two different countries (45 in Spain and 45 in Portugal), voluntarily participated in a listening survey, defined in close agreement with the objectives already stated. The ages of the respondents ranged from 18 to 61 years old and gender-balance rules were observed. The sample of participants on this study include mainly university students, doctoral students, university professors and professionals from other sectors.

### 2.2. Noise stimuli

#### 2.2.1 Traffic noise spectra and noise attenuation filters

Since the scope of the present study involves the assessment of noise reducing devices, representative stimuli of road traffic noise had to be monitored in real conditions. Therefore, dozens of road traffic noise samples were recorded, in the vicinity of important roadways, and, at the end, a set of 3 noise samples were selected as being representative of the most interesting types of noise for this study (city or urban traffic noise and road or motorway traffic noise), as well as one particularly annoying road traffic noise that was also included, corresponding to a motorbike noise sample. On the other hand, one of the selected traffic noise samples was registered from vehicles in an urban environment, and the other two corresponded to vehicles passing on the motorway (separately, light vehicles and motorcycles). In Figure 2, the sound spectra of the three selected traffic noise signals (A, B and C) can be observed.

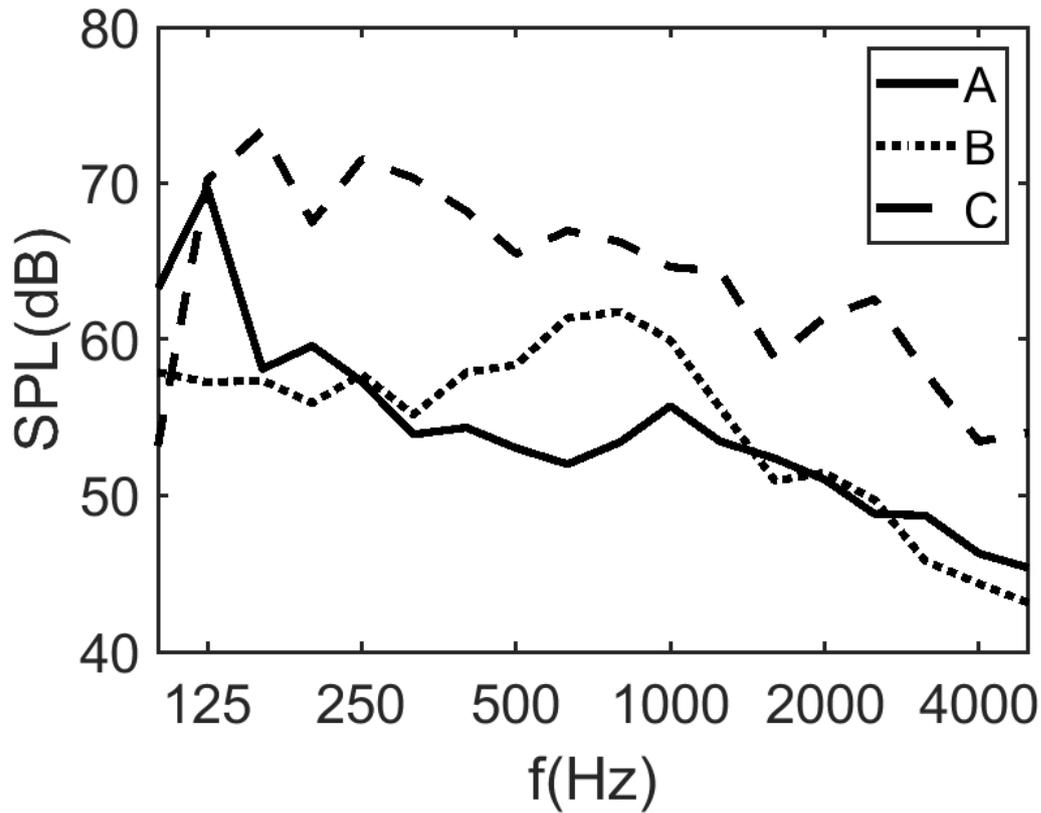

Figure 2. Sound spectra in one-third octave frequency bands of the noise traffic recordings selected for the listening tests: (A) light vehicles at urban speed; (B) light vehicles on motorways; (C) motorcycles on motorways.

|           | A    | B    | C    |
|-----------|------|------|------|
| SPL (dB)  | 71.9 | 69.5 | 79.7 |
| SPL (dBA) | 63.5 | 66.9 | 74.7 |

Table 1. Global Sound Pressure Level (SPL) values for each type of traffic sound emission selected for the study, in dB and dB(A)

These traffic sounds were chosen since they exhibit sufficiently evident spectral differences between them, as can be seen in Figure 2 and in the corresponding global Sound Pressure Levels summarized in Table 1.

Once the traffic noise signals were selected, corresponding to the sound emmissions, different sound attenuation filters were mathematically applied to each signal by post-processing the original signals. The attenuation filters represent the implementation of different noise reducing devices, and they were calculated from the data of the insertion loss values of three selected noise barrier types, with distinct characteristics, namely a classic sonic crystal noise barrier, an absorbent sonic crystal noise barrier, and a traditional continuous noise barrier. The sound attenuation filters were applied in one-third octave frequency bands, in the range between 100 and 5000 Hz, since these are the limiting bands used in the standard defining the normalized traffic noise spectrum [38].

The simulated noise barriers, regardless of their type, have always been considered with a height of 3 m. The traditional noise barrier has been considered with an estimated surface weight of 21 kg/m2, and the two sonic crystal noise barriers composed by three rows of dispersers, periodically organized in a square lattice mesh, with a regular spacing between the centers of the cylinders of 0.17 m and a filling factor of 40%.

Three different attenuation scenarios have been simulated for each noise barrier, with distinct relative positions between the noise emitter and the receiver with respect to the barrier (see schematic illustrations in Table 2 – I, II and III, represented in Figure 3, respectively, by continuous, dotted and dashed lines). For each noise barrier type, the three different situations (I, II and III) have been taken into account for the numerical evaluation of the insertion loss values along the frequency range, $IL_i$, which corresponds to a considerable number and variety of attenuation situations being simulated.

Therefore, situation I is simulating a semicircular sound wavefront being emitted from the ground level, at a distance of 2.5 m from the noise screen, and the receiver is located at a distance of 1 m from the screen and at a height of 1 m from the ground. Then, in situation II, an incident plane wave has been considered and a receiver has been simulated at a distance of 1 m from the screen and at a height of 1 m from the ground. Finally, in situation III, an emitter has been simulated as an incident plane wave and the receiver has been placed at a distance of 4.5 m from the noise barrier and at a height of 2.75 m from the ground. In short, the computed sound stimuli database includes 9 different possibilities of noise attenuation, for each selected sound signal. With all these sound stimuli, the database used to generate the attenuation filters in the listening experience has been completely and adequately generated.

| EMITTER/RECEIVER SITUATION | 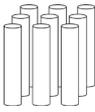 | 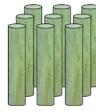 | 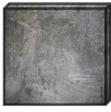 |
|---|---|---|---|
| | $DL_{SI}$ = 4,98 dB(A) | $DL_{SI}$ = 11,1 dB(A) | $DL_{SI}$ = 26,4 dB(A) |
| I 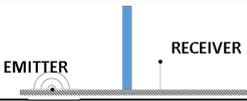 | $D_{IL,Atr}$ = 5dB(A) Sample 3 | $D_{IL,Atr}$ = 11dB(A) Sample 7 | $D_{IL,Atr}$ = 22dB(A) Sample 9 |
| II | $D_{IL,Atr}$ = 4.8dB(A) Sample 2 | $D_{IL,Atr}$ =10.5dB(A) Sample 6 | $D_{IL,A,tr}$ = 17dB(A) Sample 8 |

| | | | | | |
|---|---|---|---|---|---|
| | 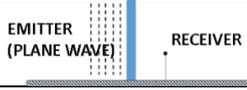 | | | | |
| III | 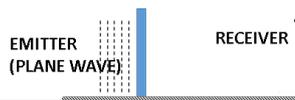 | $D_{IL,Atr}$ = 3.7dB(A)<br>Sample 1 | $D_{IL,Atr}$ = 6.5dB(A)<br>Sample 4 | $D_{ILA,tr}$ = 7.8dB(A)<br>Sample 5 | |

Table 2. Representation of noise barriers configurations, and estimated insertion loss parameters, $D_{IL,\ Atr}$, obtained from the intrinsic characteristics of the three types of noise barriers ($DL_{SI}$) and extrinsic characteristics related to the emission and reception relative positions (situations I, II and III).

In order to obtain the attenuation filters associated to each sample, and taking into account not only the intrinsic characteristics of each noise barrier but also the particular geometry and configuration (for example, noise barrier height, type of generated sound wave, position of the emitting source and receiver), a simple methodology has been adopted, based on the Huygens-Frenel-Kirchoff principle. In fact, the portion of the acoustic energy not blocked by the noise screen has been calculated numerically, using a 2D simulation model, based on a Finite Difference in the Time Domain (FDTD) method previously developed and validated by the authors [39]. Then, the two contributions at the receiver, namely the sound traveling over the noise barrier and the sound going through the noise screen, are added assuming incoherence of both contributions. One should note that the second contribution, the sound going through the noise screen, is partially attenuated in comparison to the incident sound wave. The acoustic attenuation provided by the noise barriers is estimated from the sound reduction index of each noise barrier. Further details can be found at [40].

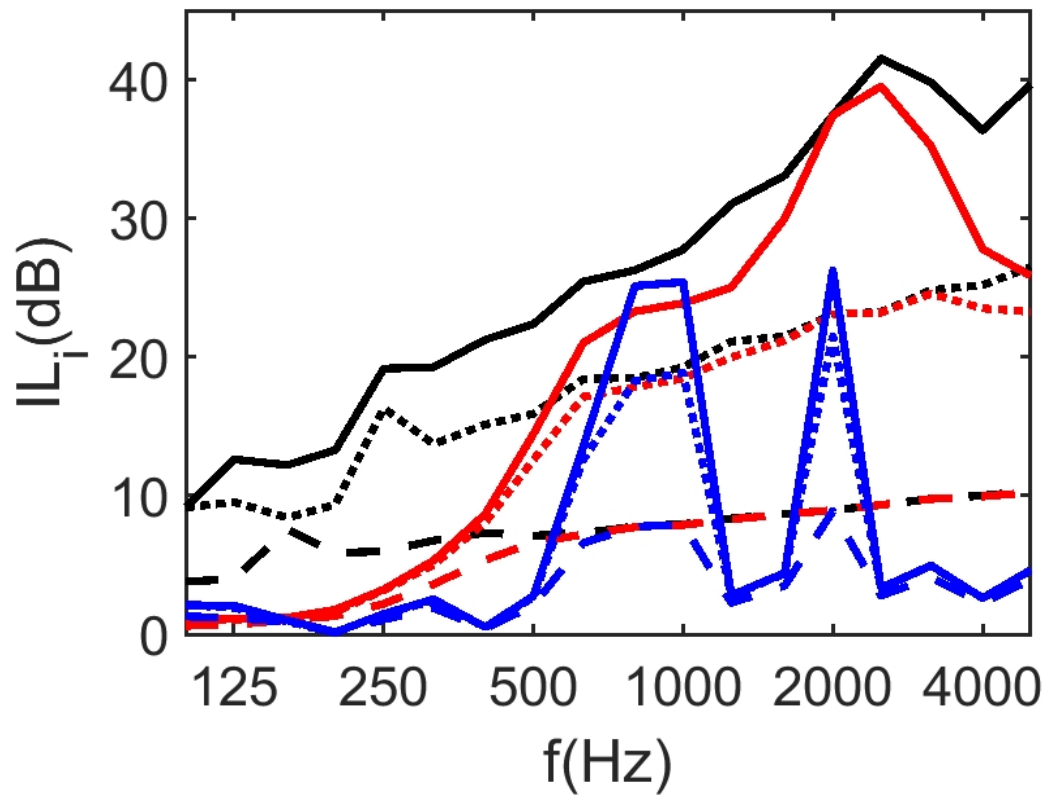

Figure 3. Estimated insertion losss levels, $IL_i$, along the analyzed frequency range, for the 9 sound attenuation samples in one-third octave bands: different situations I, II, III represented, respectively, as continuous, dotted and dashed lines; on the other hand, traditional noise barrier, sonic crystal noise barrier with absorbent scatterers and classic sonic crystal noise barrier represented in lines black, red and blue, respectively.

As it can be seen in Figure 3, the classic sonic crystal noise barrier, incorporating completely rigid and reflective scatterers, acts as very selective attenuation filters, only affecting very specific ranges of frequencies. These same crystal elements, when coated with absorbent material, leads to the amplification of their attenuation capacity to the medium and high frequencies, but not in the lower range of frequencies. In reality, only traditional continuous noise barriers can present some attenuation at low frequencies, while exhibiting increasing sound reduction values with the frequency increase.

### 2.2.2 Objective parameters characterizing noise mitigation

In this study, the sound attenuation provided by noise barriers is expressed by the airborne sound reduction single parameter, $DL_{SI}$, attending to intrinsic features of the mitigation devices. This parameter arises by taking into account the energy losses when the sound waves crosses the barrier. On the other hand, taking also into account other extrinsic features, such as the energy that is diffracted at the top of the noise barrier or the relative position of the noise emitter and receiver, the single parameter defining the noise barrier's insertion loss, $D_{ILA, tr}$, is used too.

In both cases, the information corresponding to the frequency spectrum in one-third octave bands is weighted into single value ratings, according to the following expressions.

The airborne sound insulation rating, $DL_{SI}$, has been evaluated following standard [27], being weighted with the standardized traffic noise spectrum [38]:

$$\text{Eq. (1): } DL_{SI} = -10 \log \left[ \frac{\sum_{i=m}^{18} 10^{0,1L_i} 10^{-0.1SI_i}}{\sum_{i=m}^{18} 10^{0,1L_i}} \right]$$

where:

$L_i$ is the A-weighted standard sound pressure level, in decibels, of road traffic noise within the $i^{th}$ one-third octave band of the spectrum defined in EN 1793-3 [38],

$SI_i$ is the acoustic reduction index, in the $i^{th}$ one-third octave band, of the noise barrier.

This parameter has been selected to take into account the intrinsic characteristics of the noise barriers being analyzed.

Additionally, in order to quantify the acoustic performance, accounting for the different extrinsic characteristics of each noise barrier, the single rating parameter $D_{ILA,tr}$ has also been calculated, based on standards [27, 28], ensuring that both parameters are equally weighted (Table 3).

$$\text{Eq. (2): } D_{IL,Atr} = -10 \log \frac{\sum_{i=1}^{18} 10^{0,1L_i} 10^{-0,1IL_i}}{\sum_{i=1}^{18} 10^{0,1L_i}}$$

where:

$IL_i$ is Insertion Loss of the noise barrier, in the $i^{th}$ one-third octave band.

Therefore, in both cases, the noise barriers sound attenuation ratings were weighted by the A-weighting curve and the normalized traffic noise spectrum [38] and evaluated by single-number indices as already mentioned (Table 3). The computed single-number ratings characterizing the three types of noise barriers and the three situations being analyzed are presented in Table 2, ranging from 3.7 dB(A), for the classic sonic crystal noise barrier under situation III, to a maximum attenuation performance of 22 dB(A), corresponding to the traditional noise barrier under situation I.

| | | |
|---|---|---|
| $SI_i$ | $IL_i$ | Frequency parameters. One value for each $i^{th}$ one-third octave band. |
| $DL_{SI}$ | $D_{IL,Atr}$ | Single-number ratings, weighted by the A-weighting curve and the normalized traffic noise spectrum. |

Table 3. Objective parameters used in the study

## 2.3. Listening survey procedure

The present study was developed in two South European countries and therefore in two different sites. In order to ensure analogous environments in both countries for performing the listening tests, the use of a dedicated headphone system in the survey, connected to a laptop computer from which the sound emission was controlled, was considered to be more appropriate for the accuracy of the responses. The conditions and environments where the listening tests took place were controlled to ensure that the conditions during all experiments were similar. In fact, working with the same laptop and a dedicated headphone system ensured that the conditions were equivalent, given that the same devices were used in both cases.

An application was designed and implemented in MATLAB R2019 to control and perform the listening survey. It allowed for the clear presentation of the purpose of the survey and operational instructions, then, the ordered emission and listening of the noise signals followed. At the same time, the answers given to the listening tests have been collected and successfully stored. Some degree of versatility was given to the participants, being able to repeat each played signal before answering or moving to the next sound. The sound events were sorted and played randomly.

The computer application was designed to allow the rating of perceived attenuation provided by the different noise barriers, based on six possible responses given by the listening test respondents. Therefore, when each test respondent was asked about his Perceived Annoyance Reduction (*PAR*), compared to the original sound (the emitted traffic noise without attenuation). Below you can see the possible answers that the user could choose to evaluate the PAR value in relation to the numerical value that has subsequently been used in the analysis of results (Nothing=0. Very Little=2. Little=4. Enough=6 Much=8. A lot=10.)

Since the survey participants were either Portuguese or Spanish native speakers, the application was designed in English language to serve all participants equally and avoid misinterpretations related to the question formulation.

The procedure followed during each listening survey carried out is described below. The different steps that have been followed with each interviewed individual are here detailed:

1. First of all, an audiometric test was performed, to rule out individuals with any type of hearing defect.

2. After the audiometric test, the procedure of the listening survey was explained. A presentation was used to ensure that all test respondents received the same information.

3. The listening survey began when the individual was comfortably installed and the dedicated headphone systems correctly put on. The noise signals were reproduced, for the first time they were played, in pairs (i.e the unfiltered traffic noise sample followed by the noise attenuated by the noise reducing device) so that the test listeners had to select one of the above mentioned options to evaluate the effectiveness of the device, before moving on to the next sound event.

The complete listening test (audiometry, presentation with explanation and noise reduction listening survey) did not take longer than approximately 20 minutes, depending on the number of repetitions of the same sound event that was selected by the respondent.

Respondents were exposed to four series of ten pairs of sounds (first, the unfiltered noise and, then, the attenuated noise). Therefore, each test respondent listened to a total of 40 pairs of noise events. Firstly, they heard all the attenuation combinations referred to noise signal A, then those to noise signal B, and then those to noise signal C. Finally, they listened for a second time to the noise signal of A series. The first series of noise signal A was not registered because the respondents needed some training time to adapt to the listening survey and the computer application they were using for responding.

An analysis of the response times was carried out to check that the familiarization and training phase was long enough, and the average answering delays from all respondents can be observed in Figure 4. The evident stabilization in the average response time, approximately after the first group of 10 questions, seems to indicate that the first part of the listening test, with the first hearing of noise signals A, has been sufficient to adequately train the survey respondents.

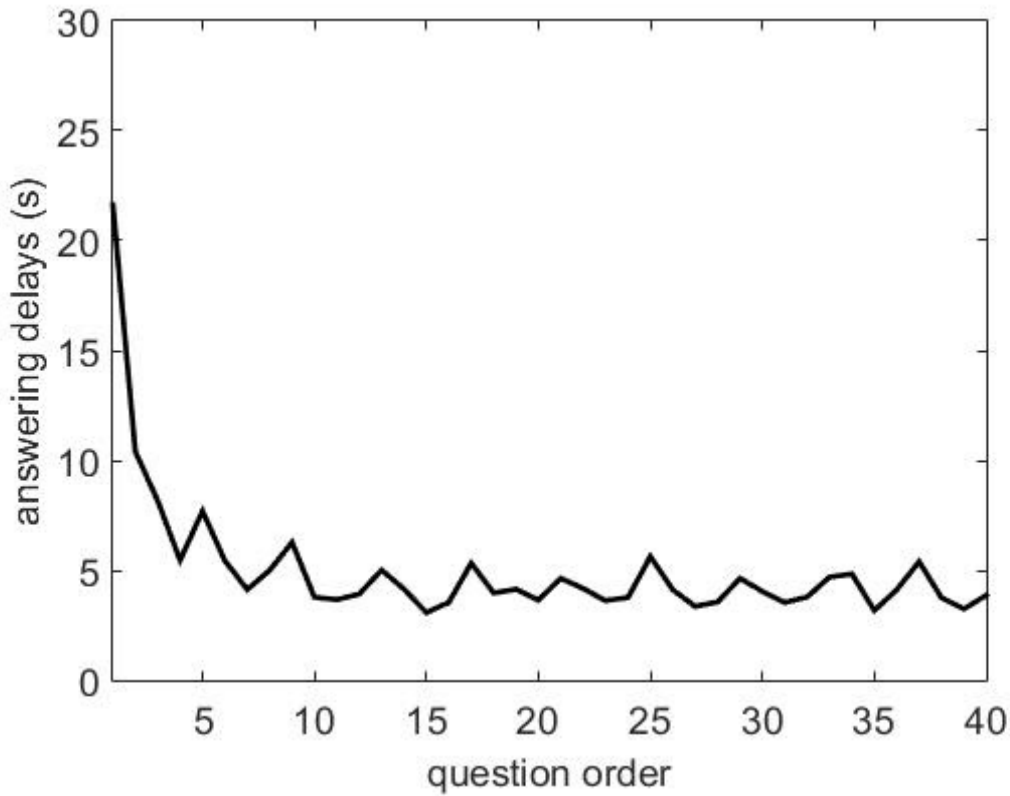

Figure 4. Average response time of the 40 questions asked.

Within each series of noise signals, as mentioned above, the noise events were played in random order, but always starting with the most attenuated noise signal, i.e. the one with the highest $D_{IL,Atr}$ value. Accordingly, each survey respondent was informed of this circumstance, allowing this reference noise signal to be taken by the listener as an "anchor" of the survey within each series noise.

**2.4. Analysis methodology**

Once the database was compiled, a statistical analysis of the collected data was carried out. For this purpose, the SPSS statistics software was used, enabling the application of different known statistical evaluations, in two separate phases, as briefly described on Table 5 (the reader who wishes to explore the use of the software and the set of applied statistical techniques can consult Field 2005 [41]).

| Phase | Techniques | Expected result |
| --- | --- | --- |

| | Correlation between objective parameters and subjective assessment | - Spearman correlation | Correlation between objective parameters ($D_{IL,Atr}$ and $DL_{SI}$) and subjective assessment (*PAR*) |
|---|---|---|---|
| a | | | Correlation between objective parameters ($D_{IL,Atr}$ and $DL_{SI}$) and subjective assessment (*PAR*), differentiating by noise type |
| | | | Correlation between objective parameters ($D_{IL,Atr}$ and $DL_{SI}$) and subjective assessment (*PAR*), differentiating by respondent characteristics |
| b | Analysis of significant differences in assessment | - Mean analysis<br>- Kruskal-Wallis test/ Mann-Whitney test<br>- Mean analysis<br>- Friedman test and Wilcoxon's post-hoc analysis | Significant differences in subjective assessment (*PAR*), according to the age, gender and country of the respondent |
| | | | Significant differences in subjective assessment (*PAR*), according to noise type |

Table 4. Data treatment phases, statistical techniques and expected results.

## 3. Results of the listening test

The analysis of the data collected from the listening survey followed the methodology listed in summary in Table 4. Firstly, the results regarding the main objective of this work are presented, corresponding to the relationship between objective parameters used for noise barriers characterization and the subjective assessment perceived by the respondents to the listening tests. Secondly, significant differences observed in the subjective assessment, detected in the Perceived Annoyance Reduction (PAR) responses, are analysed.

### 3.1 Correlation between objective parameters ($D_{IL,Atr}$ and $DL_{SI}$) and subjective assessment (*PAR*)

In this section, the correlation between the objective indicators measuring the acoustic performance of noise reducing devices and the respondents' subjective assessment is analyzed, obtaining a first general correlation, and then distinguishing between the acoustic characteristics and the characteristics of the respondents themselves.

The statistical treatment for this analysis depended on the normality of the data for each variable. Kolmogorov-Smirnov (K-S) test were used to examine the normality of data.

The data corresponding to the objective single-number indicators ($D_{IL,A,tr}$ and $DL_{SI}$) and the subjective assessment variable (PAR) follow a non-normal distribution (K-S test, p<0.05), so the Spearman rank-order correlation (non-parametric test) is used.

### 3.1.1 Correlation between $D_{IL,Atr}$ and $DL_{SI}$ and subjective assessment (*PAR*)

The Spearman correlation coefficient, significant at p < 0.05, between the subjective assessment variable (PAR) and the objective indicators ($D_{IL,A.tr}$ and $DL_{SI}$) are analysed, respectively, in Figure 5 and Figure 6, taking into account all the answers by the test respondents. A high value of the Spearman's correlation coefficient (rho), approaching +1, and a significance level of 0.000, indicates significant and very high correlations with both indicators, especially in Figure 5 with $D_{IL,A,tr}$. However, it can also be observed in Figure 6 that, as the values related to the objective sound attenuation parameter ($DL_{SI}$) increase, the subjective perceived response does not grow proportionally, tending towards a saturation value in the annoyance reduction perception value (PAR).

| Spearman's rho | $D_{IL,Atr}$-PAR |
|---|---|
| Correlation coefficient | 0.812 |
| Sig. | 0.000 |

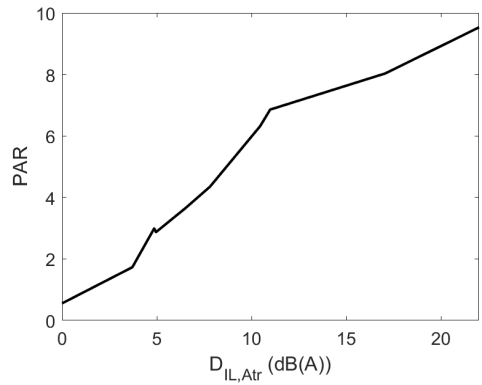

Figure 5. Correlation between $D_{IL, Atr}$ objective parameter and subjective assessment, *PAR*.

| Spearman's rho | $DL_{SI}$-PAR |
|---|---|
| Correlation coefficient | 0.695 |
| Sig. | 0.000 |

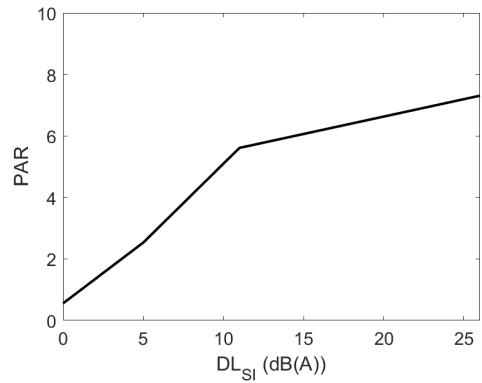

Figure 6. Correlation between $DL_{SI}$ objective parameter and subjective assessment, PAR.

### 3.1.2 Correlation between $D_{IL,Atr}$ and $DL_{SI}$ and subjective assessment (PAR), differentiating by noise type

When separated by noise type (A, B and C, as described in Section 2.2), the interesting level of correlation is still maintained (Figure 7 and Figure 8). Thus, with all three types of noise the correlations between the target indicators ($D_{ILAtr}$ and $DL_{SI}$) and PAR subjective assessment are significant (p<0.05) and with a high correlation coefficient (above 0.82). Analyzing both figures 7 and 8, it becomes evident that the test respodents perceive motorcycle traffic noise in a clearly differentiated way from that of light vehicles, either in urban areas or in motorways. In fact, the subjective annoyance perception of the motorbike traffic noise generates exhibit lower values of PAR, independently of the type of noise barrier being used to mitigate traffic noise. Once again, the behavior observed by the correlations of Figure 8 illustrate the presence of a saturation value when dealing with the single-number rating related to the insertion loss provided by the noise barriers ($D_{ILAtr}$).

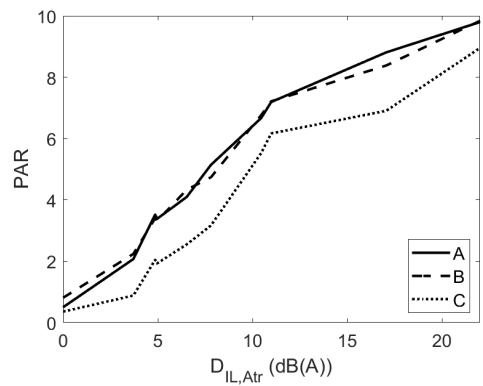

Figure 7. Correlation between $D_{IL,Atr}$ objective parameter and *PAR*, separating by noise type.

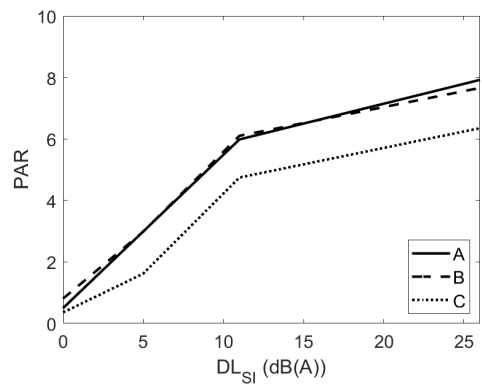

Figure 8. Correlation between $DL_{SI}$ objective parameter and *PAR,* separating by noise type

### 3.1.3 Correlation between $D_{IL,Atr}$ and $DL_{SI}$ and subjective assessment (*PAR*), differentiating by respondent characteristics

Two respondent characteristics can now be considered, namely the country where each part of the listening test took place and the gender of the test respondents. Very high and significant correlations ($p<0.05$) are observed when separating the analysis of results by country, with higher Spearman's correlation coefficients in the case of Portugal. The Portuguese individuals that participated in the listening test have perceived the changes in the annoyance reduction with slightly greater intensity than the Spanish respondents (Figure 9 and Figure 10), in terms of the effectiveness of the noise barriers, independently of the type of the noise barrier considered. This could be an interesting result, taking into account the differences between the regulations related to environmental noise control between the two countries.

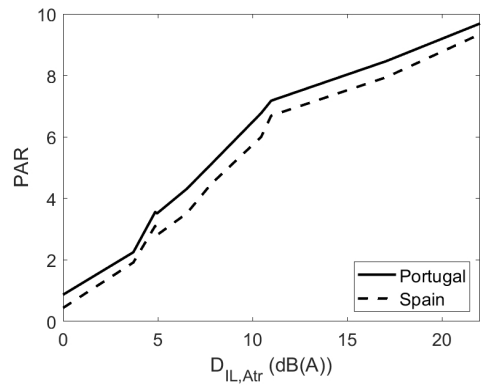

| Spearman's rho | COUNTRY ($D_{IL,Atr}$-PAR) | |
|---|---|---|
| | Spain | Portugal |
| Correlation coefficient | 0.807 | 0.823 |
| Sig. | 0.000 | 0.000 |

Figure 9. Correlation between $D_{IL,Atr}$ objective parameter and *PAR*, separating by country.

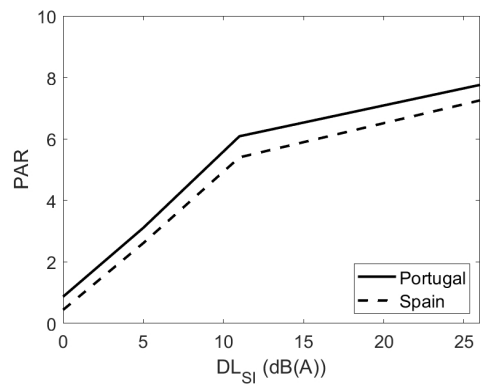

| Spearman's rho | COUNTRY ($DL_{SI}$-PAR) | |
|---|---|---|
| | Spain | Portugal |
| Correlation coefficient | 0.689 | 0.707 |
| Sig. | 0.000 | 0.000 |

Figure 10. Correlation between $DL_{SI}$ objective parameter and *PAR*, separating by country.

Performing the statistical analysis while separating the listening test sample between men and women respondents, a strong correlation is still maintained (cf. Figure 11 and Figure 12). In general, women have higher Spearman's coefficient correlations with both objective indicators, $D_{IL,Atr}$ and $DL_{SI}$. The graphic representations demonstrate that the female respondents are more sensitive to noise or, in other words, that there is less reduction in the perception of noise annoyance by women, in any of the situations herein studied.

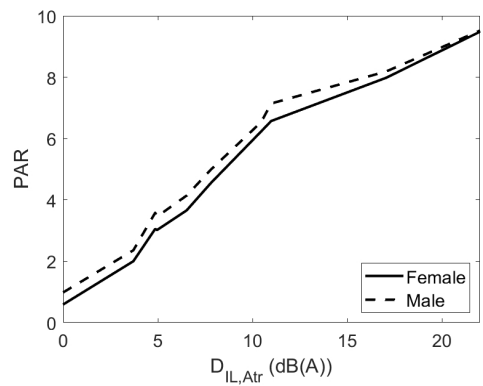

Figure 11. Correlation between $D_{IL,Atr}$ objective parameter and PAR, separating by gender.

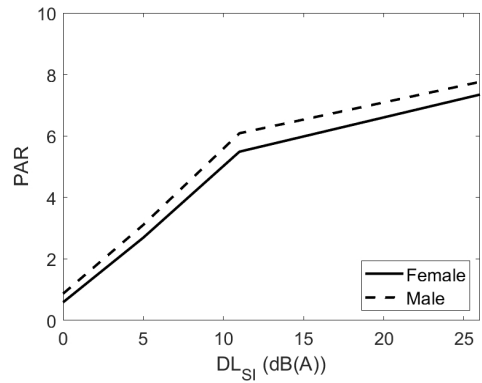

Figure 12. Correlation between $DL_{SI}$ objective parameter and PAR, separating by gender.

The analysis of the collected data, differentiating by the age of the respondents is not here presented because, as it will be seen in in Section 3.2, there are no significant differences in the subjective parameter PAR, depending on the age of the respondents (p=0.113).

**3.2 Analysis of significant differences in assessment**

In this section, possible significant differences in the Perceived Annoyance Reduction assessment (PAR) by all participants are now analyzed according to their own characteristics (for instance, age, gender and country) and the acoustic characteristics of the traffic noise being mitigated by the devices (the three noise types).

Since the collected data representing subjective perception (*PAR*) do not follow a normal distribution (K-S test, p<0.05), different non-parametric tests are adopted for the following statistical analyses: Mann-Whitney and Kruskal-Wallis tests, respectively comparing 2 (for country or gender characteristics) or k (for respondents age) independent samples; and Friedman test comparing k (noise type) dependent samples).

**3.2.1 Significant differences in subjective assessment *(PAR)*, according to the age, gender and country of the respondents**

Through the application of the Kruskal-Wallis test it is possible to verify that there are no significant differences in the subjective evaluation *PAR* depending on age of the respondents (Chi square=5.964; df=3; p=0.113). Therefore, and since the population responds homogeneously in terms of age, this variable does not segment the sample in the rest of the analysis.

On the other hand, the application of the Mann-Whitney test highlights significant differences in the degree of the perceived annoyance reduction, depending on the respondent gender (Mann-Whitney U=488492.500; p=0.000) and country (Mann-Whitney U=574146.000; p=0.000). In terms of gender, Figure 13 shows that men present higher *PAR* values than women. With regard to the country, the subjective evaluations of the listening tests in Portugal, in relation to *PAR*, correspond to slightly higher values than those carried out in Spain.

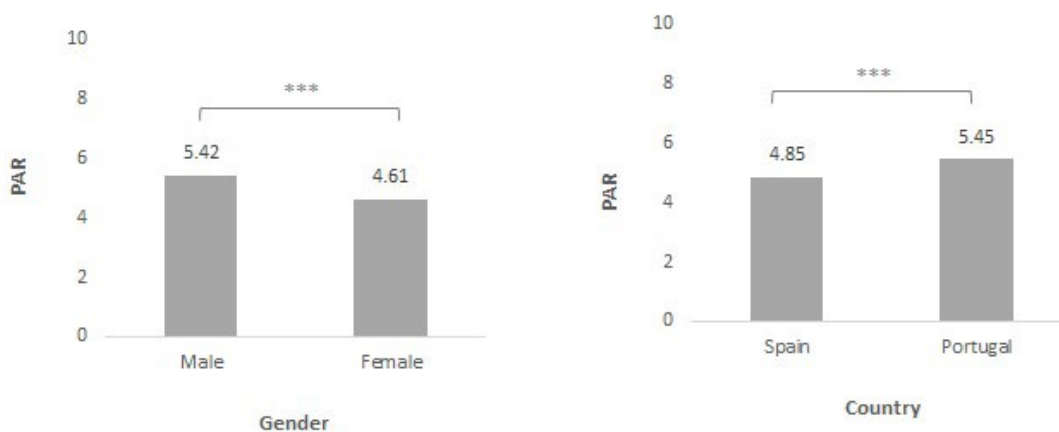

Figure 13. Average levels of *PAR* by gender and by country. Keys indicate the comparisons and asterisks the significance level (*p < 0.05, **p < 0.01, ***p < 0.001).

### 3.3.2. Significant differences in subjective assessment *(PAR)*, according to the emitted traffic noise type

The application of the non-parametric Friedman statistical test indicates that there are significant differences in the subjective evaluation *PAR*, depending on the type of the traffic noise emitted (Chi square= 385. 334; df=2; p=0.000). By performing a post-hoc analysis with Wilcoxon signed-rank tests, important correlations are observed between motorbike traffic noise and the other two types of traffic noise, urban traffic noise (p=0.000) and motorway traffic noise (p=0.000). In Figure 14, a lower level of the subjective evaluation *PAR* can be noticed, in the case of motorbike induced traffic noise, in comparison to *PAR* values for light vehicles both in urban and motorway environments.

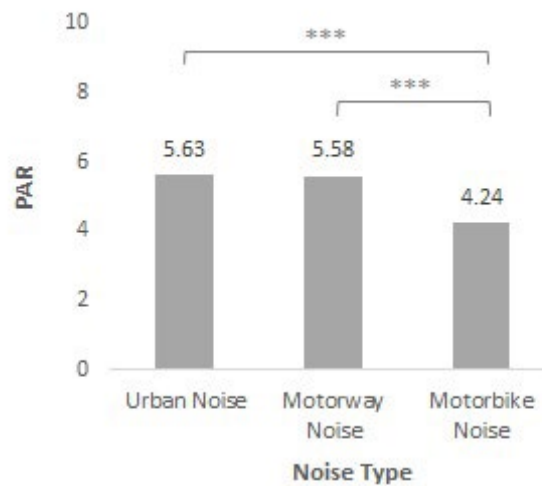

Figure 14. Average levels of *PAR* by emitted traffic noise type. Keys indicate the comparisons and asterisks the significance level (*p < 0.05, **p < 0.01, ***p < 0.001).

Finally, when separating the responses to the listening test by gender and by country, an analogous situation can be observed, with higher subjective *PAR* ratings for male respondents and Portuguese tests (Figure 15). The application of the non-parametric Friedman statistical test enables detecting significant differences in *PAR* values, depending on the type of emitted traffic noise in male (Chi square=132.831; df=2; p=0.000) and female (Chi square=251.034; df=2; p=0.000) test respondents, and in the Spanish (Chi square=227.752; df=2; p=0.000) and Portuguese (Chi-square=160.018; df=2; p=0.000) parts of the listening test performed. Additionally, by using post-hoc analyses with Wilcoxon signed-rank tests, it is possible to see that, in all cases, these differences occur once again between the motorbike traffic noise and the other two types of emitted traffic noise.

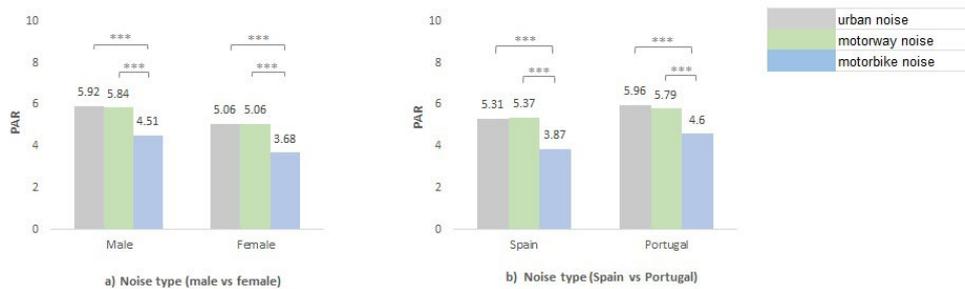

Figure 15. Average levels of *PAR* by emitted traffic noise type, according to gender and country. Keys indicate the comparisons and asterisks the significance level (*$p < 0.05$, **$p < 0.01$, ***$p < 0.001$).

## 4. Discussion

Different research studies have been developed on the effectiveness of some devices that provide acoustic insulation, both in terms of airborne sound reduction and impact sound insulation in building acoustics context [35, 36, 32, 43]. Also, some perception studies have been carried out on the acoustic comfort provided by noise reducing devices installed close to transport infrastructures (roads and railways) [44, 45], where either acoustic performance of noise barriers and their relation with the subjective annoyance reduction were studied, or predictions of annoyance and exposure-response curves (in $L_{den}$) were compared searching for significant correspondences.

However, the adequacy of the objective parameters used with respect to the annoyance in environmental noise situations has not been sufficiently studied. The effects of audio-visual variations on the perception of the acoustic performance of noise barriers have been considered by Hong and Jeon [37], but not its correlation with objective parameters. In fact, technical interventions for reducing noise levels may not lead to proportional impacts on annoyance reduction [46], according to the objective indices with which they are assessed. Therefore, all of this justifies the interest of the present work.

Currently, in order to evaluate the acoustic performance of a noise barrier, the most frequently used target indices are the above described intrinsic parameters, $DL_R$ and

$DL_{SI}$, depending whether they are measured on diffuse sound field or direct sound field conditions. It is common to use only the single number $DL_R$ or $DL_{SI}$ to determine the quality of a noise barrier. In fact, some recent technical publications even establish requirements of a minimum sound insulation ($DL_{SI}$) value of 28 dB [47]. But, as we have seen, there is another single number index that considers, not only the intrinsic conditions of the noise barrier itself, but also the extrinsic conditions of the implemented noise barrier and the measurement environment, representing an Insertion Loss rating, standardized in ISO 10847 [28]. Based on this standard, a single-number quantity for rating the insertion loss weighted by standardized traffic noise spectrum has been defined, $D_{IL,A,tr}$.

In the present study, the correlation between these objective indices that characterize the acoustic barriers ($DL_{SI}$ and $D_{IL,Atr}$) and the perception of annoyance reduction (*PAR*) felt by respondents of a listening test is analyzed. In fact, a high correlation between these objective and subjective indices has been observed.

Previous studies have shown that the configuration of the urban environment determines the effectiveness of noise barriers, thus demonstrating the importance of their extrinsic characteristics [48]. In the present work, it can be verified that, indeed, there is a better correlation between the subjective evaluation *PAR* and the objective index which considers the insertion loss achieved by the noise barrier and its extrinsic characteristics ($D_{IL,Atr}$) than the single-number rating that considers intrinsic characteristics as the airborne sound reduction ($DL_{SI}$).

Another interesting aspect extracted from the described results is that, for both objective indices, the subjective evaluation *PAR* is getting saturated above certain values of attenuation; in other words, the increase of attenuation provided by the noise reduction devices, does not lead to a relevant increase of the subjective reduction *PAR*. In fact, above certain values of airborne noise reduction, the perceived acoustic comfort hardly improves at all. In 1968, Maekawa [49] already pointed out that it was not possible to achieve noise barrier attenuation above a certain value. This limit, verified experimentally, depends on the particular thickness of the noise barrier and ranges between 20 and 25 dB. The results of the present psychoacoustic test confirm this conclusion. This is an important factor to bear in mind when designing noise reduction devices, since efforts should not focus on increasing attenuation performance above certain values, but once these values are reached, research efforts should better be focused on obtaining lighter devices, that require less foundations, smaller areas occupied at the sides of the infrastructures and better acceptance by citizens.

Concerning secondary objectives of the study, although previous studies have found that demographic variables such as age, gender and type of housing are unimportant noise annoyance modifiers in steady state noise conditions [50, 51], after the analysis of the results with respect to the characteristics of the test respondents in the present study it can be seen that the correlation between subjective and objective data is present in all cases, but there are significant differences with respect to the gender and nationality of those surveyed. Thus, comparing the results between men and women, female respondents were more demanding than male respondents when it comes to evaluating acoustic comfort. The same happens in the case of Spanish respondents as compared to Portuguese ones, being the Spaniards more demanding in their answers than the Portuguese. Then, in view of the results, gender and nationality are relevant factors to consider when evaluating the effectiveness of noise reduction devices.

Significant differences were also observed with respect to the traffic noise used as emission source. The level of correlation is very similar in all three cases; however, the motorbike traffic noise has been shown to have a smaller reduction in annoyance perception than the rest. These differences in the perceived annoyance reduction as a function of emitted and attenuated noise are in accordance with what has been found in previous studies [42].

The results obtained in this work are very interesting, especially from the point of view of non-proportionality and the observed trend to saturation of the perceived reduction of traffic noise annoyance. This should encourage further studies aimed at finding the best noise barrier characterization index and clearly verifying what attenuation limits may exist.

The data analysed in this paper enables the identification of an interesting trend, that requires further research in order to get its consolidation, when considering different frequency weightings from the A-weighting curve. In Figure 16, the results already presented in Figure 5 are now being compared to alternative results obtained by replacing the A-weighting with other frequency weighting curves. Bear in mind that the weighting curves A, B, C, and others, were established to be use at different average sound levels. In fact, the A-weighting curve was originally introduced to deal with lower sound levels (around 40 phon). On the other hand, the B-weighting curve corresponds approximately to 60 phon, and it is appropriate to deal with intermediate sound levels. In addition, we have also computed and introduced in the present analysis the -weighting curve inspired in the equal loudness of 10 phon, which has been designated as "alpha" in Figure 16, allowing for the representation of very low sound levels. Attention should also be drawn to the fact that the A-weighting curve is nowadays widely used due to its good correlation between the pollution measurements and industrial noise, in relation to the occupational deafness and human hearing annoyance. Therefore, Figure 16 illustrates that the higher the loudness level considered in phon, the relationship between both objective and subjective indices represented in this figure become less linear. It should be noted that the main difference between the "alpha", A and B-weightings is the relevance level of the low frequencies sound components. This observed trend seems to indicate that the usual procedures could be overestimating the relevance of low frequencies in the evaluation of noise barriers for reducing road traffic noise. Further research in this area will be necessary to reach conclusions.

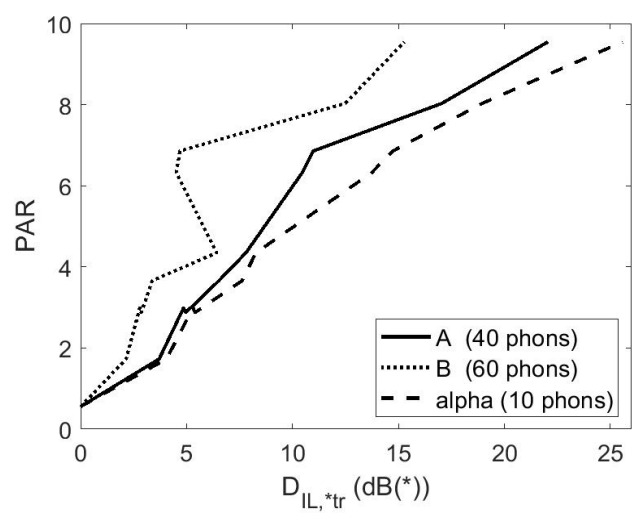

Figure 16. Perceived annoyance reduction (PAR) versus three different possible frequency weighting curves applied for computing objective indices $D_{IL,*tr}$

## 5. Conclusions

Results have been presented regarding a psychoacoustic study where the response of a group of individuals has been studied in relation to the reduction of perceived sound annoyance when traffic noise is attenuated by using a noise barrier. The study was based on sound stimuli using traffic noise, with two stimuli related to light vehicles (driving in urban areas and on motorways) and one related to motorcycles traffic noise emission. On the other hand, these sound stimuli have been filtered taking into account 9 possible attenuations of greater or lesser objective value.

Two particularly interesting results have been obtained from the analysis of the data obtained. Firstly, it is observed that the reduction of perceived annoyance does not increase proportionally to the attenuation of the noise reducing device. A relatively proportional growth is observed in the range of the first 10 dB of noise attenuation, but from that point onwards the slope grows less and with a certain trend to a saturation level. This seems to indicate that the improvement in noise attenuation, estimated by the introduction of a noise barrier has an upper limit value.

On the other hand, significant differences have been noticed between the opinion of individuals consulted on the basis of their gender and place of origin, observing that, in general, female respondents are more sensitive to noise, or, in other words, the level of perceived annoyance reduction are always lower than those of male respondents. Regarding the place of listening test accomplishment, a greater level of sensitivity corresponds to the sample of respondents from Portugal, i.e., the Spaniards seem to detect a greater reduction in perceived noise annoyance, which could possibly indicate a lower hearing sensitivity. Besides the interest of the present research study, more listening tests should be carried out taking into account these significant differences and expanding the database of noise attenuation filters in order to more accurately assess the maximum perceived reduction saturation values

Additionally, it shall be also mentioned that the presence of a saturation limit on the perceived annoyance reduction provided by noise barriers could be very relevant if other objectives are included in the design process, widening the study to supplemental characteristics, such as, increasing fluid permeability to wind, reducing visual impact, minimizing space occupied by the noise barriers, reducing the foundation structural requirements or improving aesthetics in the design of innovative noise barriers. In this case, the use of multi-objective evolutionary optimization techniques can be applied. A number of published works already propose the use of such techniques, to increase the performance of sonic crystals noise reducing devices, mainly due to the simplicity of the genetic algorithmic codification [11, 52, 53]. The presence of the above mentioned saturation limit would lead to a convex Pareto front that makes search for compromise solutions with good performance while attending to all cost functions easier.

Both acoustic and landscape issues need to be taken into consideration to design effective noise barriers in urban environments [37]. For this reason, following these conclusions, further research could be carrying out of audiovisual perception tests to measure the effective performance of traditional barriers and SCAS.

**Acknowledgments**

This work has been financed by national funds through FCT – Foundation for Science and Technology, I.P., within the scope of the R&D unit Institute for Sustainability and Innovation in Structural Engineering - ISISE (UIDP/04029/2020) and through the Regional Operational Programme CENTRO2020 within the scope of the project CENTRO-01-0145-FEDER-000006 (SUSpENsE).

Also this work has been supported by the Ministerio de Ciencia, Innovación y Universidades, Spain, under grant RTI2018-096904-B-I00

M.P.P.T. is grateful for the support of pre-doctoral Grant by the Ministerio de Economía y Competitividad of Spain through reference No. DI-15-08100.

*'Declarations of interest: none'*